\documentstyle[12pt]{article}

\def\z{\hspace*{9mm}}
\def\x{\hspace{3mm}}
\def\ar{\begin{array}{rcl}}
\def\an{\end{array}}

\newcommand{\eq}{\begin{equation}}
\newcommand{\eqa}{\begin{eqnarray}}
\newcommand{\en}{\end{equation}}
\newcommand{\ena}{\end{eqnarray}}
\def\1{{\bf 1}}

\def\ot{\otimes}
\def\id{\mbox{id}}
\def\up{\uparrow}
\def\idA{{\bf 1}_{\cal A}}
\def\g{\mbox{\bf g\,}}
\def\C{{\cal C}}
\def\uqg{\mbox{$U_{q}{\/\mbox{\bf g}}$ }}
\newcommand{\tr}{\triangleright}
\newcommand{\trc}{{\,\stackrel{c}{\triangleright}\,}}
\def\A{\mbox{${\cal A}_{\pm}$}}
\def\R{\mbox{$\cal R$\,}}
\def\F{\mbox{$\cal F$}}
\def\Fu{{\cal F}^{(1)}}
\def\Fd{{\cal F}^{(2)}}

\def\ie{\mbox{\it i.e.\/ }}
\def\eg{\mbox{\it e.g.\/ }}
\newcommand{\cn}{{\bf C}}
\newcommand{\rn}{{\bf R}}

\newtheorem{prop}{Proposition}
\newtheorem{lemma}{Lemma}
\newtheorem{theorem}{Theorem}

\begin{document}
\begin{titlepage}
\begin{center}
July 1996, rev. Nov. 1996          \hfill       LMU-TPW 96-17\\
\vskip.6in

{\Large \bf On Bose-Fermi Statistics, Quantum Group Symmetry, and
Second Quantization}\footnote{Talk presented at the XXI International 
Colloquium on Group Theoretical Methods in Physics (Group21), 
15 - 20 July 1996, Goslar, Germany. To appear in the proceedings
of the Quantum Group volume of Group21, Eds.  H.-D. Doebner and V.K. Dobrev.}
\footnote{Work supported by the European 
Community, through the European Commission, Dir. Gen. for Science,
Research and Development, providing the TMR grant ERBFMBICT960921.}

\vskip.4in

Gaetano Fiore

\vskip.25in

{\em Sektion Physik der Ludwig-Maximilians-Universit\"at
M\"unchen\\
Theoretische Physik --- Lehrstuhl Professor Wess\\
Theresienstra\ss e 37, 80333 M\"unchen\\
Federal Republic of Germany}

{\footnotesize 
{\it e-mail: }Gaetano.Fiore \ @ \ physik.uni-muenchen.de}
\end{center}

\vskip1in
 
\begin{abstract}
Can one represent quantum group covariant $q$-commuting
``creators, annihilators'' $A^+_i,A^j$ as operators
acting on {\it standard} bosonic/fermionic Fock spaces?
We briefly address this general problem and show that
the answer is positive (at least) in some simplest cases. 
\end{abstract}

\end{titlepage}

\section{Introduction}

In recent years the idea of Quantum Field Theories
(QFT) endowed with Quantum Group \cite{dr2}
symmetries has attracted considerable interest
and has been investigated especially in 2D field 
theories, in connection with socalled anyonic
statistics (when the deformation parameter
$q$ is a root of unity).
Its application to QFT in higher (\eg 3+1) space-time 
dimensions relies, among other things,
on the condition that Bose and
Fermi statistics are compatible  with 
quantum group-symmetry transformations, at
other (in particular real) values of $q$. The latter issue
in fact involves two different problems, one in
first quantized quantum mechanics 
and the other in QFT.

The first problem essentially is whether a Hilbert space
can carry both a completely (anti)symmetric 
representation of the symmetric group $S_n$ (so that
it can describe the states of $n$ bosons/fermions)
and of a quasitriangular non-cocommutative 
$*$-Hopf algebra $H$. Contrary to a quite widespread 
prejudice, we showed in Ref. \cite{fioschu1} that this is 
possible whenever $H$ can be obtained from the
universal enveloping $Ug$ of a Lie algebra $g$ by a
unitary ``Drinfel'd twist'' $\F$ \cite{dr1,dr3}.
Only, we need to 
describe the system of $n$ bosons/fermions in an unusual picture,
that is related to the standard one [involving (anti)symmetric 
wavefunctions and symmetric operators] by a
unitary transformation $F_{12\ldots n}$
{\it not} symmetric under tensor factor permutations;
$F_{12\ldots n}$ is derived from 
$\F$. The relevant point here is that even
in this scheme second quantization naturally leads \cite{fioschu2} 
to creation and annihilation operators $A^{+c}_i,A^j_c$ 
satisfying the {\it canonical} (anti)commutation 
relations (CCR), and to the standard bosonic/fermionic
Fock space representations, exactly as in the standard treatment
of second quantization. 

The second problem, which we briefly address here,
is whether nonetheless one can represent, as operators acting
on standard bosonic/fermionic spaces, algebraic objects
$A^+_i,A^j$: (1) transforming as conjugate tensors 
under the action of $H$; (2) satisfying the 
{\it quantum}, \ie $H$-covariant, 
commutation relations (QCR) \cite{puwo,wezu}. 
We report here results (proved in Ref. \cite{fio})
which allow a positive answer to requirement (1), under the
same assumption as above, and a positive answer to requirement (2):
a) in the simpler case that  $H$ is triangular; b) in the particular
case that $H=U_qsu(2)$ and $A^{+c}_i,A^j_c$ belong to the 
fundamental representation 
of $su(2)$\footnote{We expect the same result also for other
compact $U_q\g$ (at least when $\rho$ is a
fundamental representation), but so far could not prove it, due
to the limited knowledge about their $\F$.}.

We look for a realization of $A^+_i,A^j$ in the
form of formal power series in the 
$A^{+c}_i,A^j_c$\footnote{$A^{+c}_i,A^j_c$ 
tranform as tensors
under the action of the classical group,
{\it not} of the quantum group.}.
Using $\F$, in sect. \ref{qcov} we determine
a class of  candidates for $A^+_i,A^j$ fulfilling
requirement (1).  In Sect \ref{fulf} we show 
how to pick out of this class a particular set satisfying
requirement (2) under one of the assumptions a), b).
These  $A^+_i,A^j$ turn out
to be well-defined operators on the bosonic/fermion Fock space.
In sect. \ref{new} we briefly comment on the possible application
of our results to QFT.

\section{Preliminaries and notation}
\subsection{Twisting groups into quantum groups}

Let $(U\g,m,\Delta_c, \varepsilon,S_c)$ be the 
cocommutative Hopf algebra associated with the universal
enveloping (UE) algebra $U\/ \g$ of a Lie algebra \g. 
$m,\Delta_c, \varepsilon,S_c$  denote the 
multiplication comultiplication, counit and antipode
respectively; we will often drop the symbol $m$:
$m(a\ot b)\equiv ab$.

Let $\F\in U\g[[\hbar]]\ot U\g[[\hbar]]$ (we will write 
$\F=\Fu\ot \Fd$, in a Sweedler's 
notation with {\it upper} indices)  be a twist,
\ie an invertible element satisfying the relations
\eq 
(\varepsilon\ot \id)\F=\1=(\id\ot \varepsilon)\F
\label{cond2}
\en
and $\F\vert_{\hbar=0}=\1\ot\1$ ($\hbar\in\cn$ 
is the `deformation parameter').
It is well known \cite{dr1} that if \F \/ also satisfies the relation
\eq
(\F\ot \1)[(\Delta_c\ot \id)(\F)]=(\1 \ot \F)[(\id \ot \Delta_c)(\F),
\label{cond1}
\en
then one can construct a 
triangular non-cocommutative Hopf algebra 
$H=(U\g[[\hbar]],m$, $\Delta,\varepsilon,S,\R)$ 
having the same algebra
structure  $(U\g[[\hbar]],m)$, the same counit $\varepsilon$,  
comultiplication and antipode defined by 
\eq
\Delta(a)=\F\Delta_c(a)\F^{-1},\qquad\qquad\qquad
S(a)=\gamma^{-1} S_c(a)\gamma
\label{def1}
\en
(where $\gamma^{-1} := \F^{(1)}\cdot S_c\F^{(2)}$),
and (triangular) universal R-matrix
$\R:=\F_{21}\F^{-1}$ ($\F_{21}:=\Fd\ot\Fu$).
Condition (\ref{cond1}) ensures that $\Delta$ is 
coassociative as $\Delta_c$. 

Examples of \F's satisfying conditions  (\ref{cond1}),  
(\ref{cond2}) are provided \eg by the socalled `Reshetikhin
twists' $\F:= e^{\hbar \omega_{ij}h_i\wedge h_j},$
where $\{h_i\}$ is a basis in the Cartan subalgebra of \g
and $\omega_{ij}\in\cn$.

A similar result holds for genuine quantum groups.
A well-known theorem by Drinfel'd \cite{dr3} essentially 
proves, for any {\it quasitriangular}
deformation ~$H=(\uqg,m,\Delta$, $\varepsilon,S)$ 
\cite{dr2,frt}
of $U\g$, with  \g simple belonging to the classical 
A,B,C,D series,
the existence of an invertible \F ~satisfying condition 
(\ref{cond2}) such that $H$ can be obtained from $U\g$ 
through formulae
(\ref{def1}) as well, after 
identifying \uqg \/ with the isomorphical algebra $U\g[[\hbar]]$, 
$\hbar=\ln q$. 
This \F~ does not satisfy 
condition (\ref{cond1}), however the (nontrivial) coassociator
$\phi:=\F_{12,3}^{-1}\F_{12}^{-1}\F_{23}\F_{1,23}\in U\g^{\ot^3}$
still commutes with $\Delta^{(2)}_c(U\g)$, thus explaining why
$\Delta$ is coassociative in this case, too. The corresponding
universal (quasitriangular) R-matrix \R is related to 
\F ~by $\R:=\F_{21} q^{t\over 2}\F^{-1}$,
where $t:=\Delta_c(\C)-\1\ot \C-\C\ot \1$ is 
the canonical invariant element 
in $U\g\ot U\g$ ($\C$
is the quadratic Casimir).

In defining $\phi$ we have used a
 `tensor notation' 
which will be repeatedly employed in the sequel.
According to it,  
eq. (\ref{cond1}) can be rephrased as
$\F_{12}\F_{12,3}=\F_{23}\F_{1,23}$;
the comma separates the tensor factors 
{\it not} stemming from the coproduct.
On the other hand, 
we will use unbarred and barred indices to 
distinguish $\Delta$ from 
$\Delta_c$ in Sweedler's notation:
$\Delta_c(x)\equiv x_{(1)}\ot x_{(2)}$,
$\Delta(x)\equiv x_{(\bar 1)}\ot x_{(\bar 2)}$.

\subsection{Classically covariant creators and annihilators}

Let \A~ be the unital algebra  generated by $\idA$ and
elements $\{A^{+c}_i\}_{i\in I}$ and $\{A^j_c\}_{j\in I}$
satisfying the (anti)commutation relations
\eqa
[A_c^i\, , \,A_c^j]_{\pm}            &=& 0 \cr 
[A^{+c}_i\, , \, A^{+c}_j]_{\pm}  &=& 0 \cr 
[A_c^i\, , \, A^{+c}_j]_{\pm}       &=& \delta_j^i\idA
\label{ccr}
\ena
(the $\pm$ sign denotes commutators and
anticommutators respectively),
belonging respectively to some representation $\rho$ and to its 
contragradient $\rho_c^\vee = \rho^T \circ S_c$ of $U\g$ 
(${}^T$ is the transpose):
\eq
x\trc A^{+c}_i =\rho(x)^l_iA^{+c}_l   \qquad\qquad
x\trc A_c^i=\rho(S_cx)^i_lA_c^l. \qquad\qquad
\qquad x\in U\g,\x \rho(x)^i_j\in \cn. 
\label{covl}
\en
Equivalently, one says that $A^{+c}_i,A_c^i$ are ``covariant'', or
``tensors'', under $\trc$ .

\A~ is a (left) module of $(U\g,\trc)$, if the action 
$\trc$ is extended on the 
whole \A~ by means of the (cocommutative) coproduct:
\eq
x\trc (ab)=(x_{(1)}\trc a)(x_{(2)}\trc b).
\en

Setting 
\eq 
 \sigma(X):=
\rho(X)^i_jA^{+c}_i A_c^j
\label{jordan}
\en
for all $X\in \g$, one  finds that  
$\sigma: \g\rightarrow \A$
is a Lie algebra homomorphism,
so that  $\sigma$ can be extended to
all of $U\g[[\hbar]]$ as an algebra
homomorphism $\sigma: U\g[[\hbar]]\rightarrow \A[[\hbar]]$;
on the unit element we set
$\sigma(\1_{U\g}):=\idA$. 
$\sigma(X)$ commutes with the `number of particles'
$N^c:=A^{+c}_iA^i_c$. $\sigma$ can be seen
as the generalization of the Jordan-Schwinger
realization of $su(2)$,
\eq
\sigma(j_+)=A^{+c}_{\up}A_c^{\downarrow},\qquad\qquad
\sigma(j_-)=A^{+c}_{\downarrow}A_c^{\up},\qquad\qquad
\sigma(j_0)=\frac 12(A^{+c}_{\up}A_c^{\up}-
A^{+c}_{\downarrow}A_c^{\downarrow}).
\label{homo}
\en

\begin{lemma}
The (left) action $\trc:U\g\times \A\rightarrow \A$ can be 
realized in an `adjoint-like' way:
\eq
x\trc a=\sigma(x_{(1)})\: a\: \sigma(S_c x_{(2)}) ,
\qquad\qquad\qquad x\in U\g, \z a\in\A.
\label{cov}
\en 
\end{lemma}

\section{Deforming maps to quantum group covariant 
creators and annihilators}
\label{qcov}

On the other hand, it is straightforward to 
check that the definition
\eq
x\tr a := \sigma(x_{(\bar 1)}) a \sigma(S x_{(\bar 2)})
\en
allows to realize the ``quantum'' (left) action of 
$H$ on the left module $\A[[\hbar]]$,
\ie that $(xy)\tr a=x\tr(y\tr a)$
and $x\tr(ab)=(x_{(\bar 1)}\tr a)(x_{(\bar 2)}\tr b)$
$\forall x,y\in H$, $a,b\in\A[[\hbar]]$.

However, $A^{+c}_i,A_c^j$ are {\it not} covariant 
w.r.t. to $\tr$. Are there covariant
objects $A^+_i,A^j\in \A$ 
(going to $A^{+c}_i,A_c^j$ in
the limit $\hbar\rightarrow 0$)?
The answer comes from 

\begin{prop}\cite{fio}
For any invertible \g-invariant 
[\ie, commuting with $\Delta_c(U\g)$] 
elements $T,T'\in U\g[[\hbar]]\ot U\g[[\hbar]]$ the elements
\eqa
A_i^+ &:= &\sigma(Q^{(1)})A_i^{+c}
\sigma(S_cQ^{(2)}\gamma)\in\A[[\hbar]] \nonumber\\
A^i&:= &\sigma(\gamma'S_cQ^{'(2)})A^i_c
\sigma(Q^{'(1)})\in \A[[\hbar]]                        
\label{def3}
\ena
($Q:=\F T$ $Q':=\F^{-1}T'$,
$\gamma':= \left[S_c\F^{-1(2)}\cdot\F^{-1(1)}\right]^{-1}$)
are ``covariant'' under $\tr$, more precisely 
belong respectively to the irreducible representations
$\rho$ and to its {\it quantum} contragredient
one $\rho^\vee=\rho^T\circ S$ of $H$ acting 
through $\tr$:
\eq
x\tr A^+_i=\rho(x)^l_iA^+_l\z\z\z x\tr A^i=\rho(Sx)^i_mA^m.
\en 
\label{prop1}
\end{prop}
\vspace{-.5truecm}

{\bf Remark 1}. If $H$ is a $*$-Hopf algebra, $\rho$, $\F$  are
unitary\footnote{One can always choose $\F$
unitary if \g is compact \cite{jurco}.} and $\dagger$ 
is an involution in $\A$, then $\gamma'=\gamma^*$ and
\eq
(A_c^i)^{\dagger}= A^{+c}_i  \qquad \qquad \Rightarrow 
\qquad \qquad 
\sigma\circ *=\dagger\circ \sigma, \qquad 
(A^i)^{\dagger}= A^+_i.
\label{star}
\en

{\bf Remark 2}.  Let $\A^{c,inv},\A^{inv}$ be the 
subalgebras of $\A[[\hbar]]$ invariant under $\trc,\tr$
(\ie $I\in \A^{c,inv}$ iff $x\tr I=\varepsilon(x) I$,
$I\in \A^{inv}$ iff $x\tr I=\varepsilon(x) I$). It is not difficult
to prove that $\A^{c,inv}=\A^{inv}$. 
An element $I\in\A[[\hbar]]$ can be expressed as a function of
$A^i,A^+_j$ or of $A^i_c,A^{+c}_j$, 
$I=f(A^i,A^+_j)=f_c(A^i_c,A^{+c}_j)$. We will prove elsewhere
that $f=f_c$ in the triangular case, but not in the genuine
quasitriangular one. In the latter case, to a polynomial $f$
(resp. $f_c$) there corresponds a highly non-polynomial
(tipically a trascendental function) $f_c$ (resp. $f$); so the
change of generators
$A^i,A^+_j\leftrightarrow A^i_c,A^{+c}_j$ can be used to simplify
the functional dependence of $I$ (what might turn useful for 
practical purposes, \eg to solve the dynamics associated to
some Hamiltonian $I$).

{\bf Remark 3}. $A^i,A^+_j$  are well-defined as operators on the 
bosonic/fermionic Fock spaces, at least for small $\hbar$ 
[assuming that the tensors $T,T'$ are also of the form 
$\1\ot \1+O(\hbar)$]; correspondingly, the transformation
$A^i_c,A^{+c}_j\rightarrow A^i,A^+_j$ is invertible.

\section{Fulfilment of the ``quantum'' commutation relations}
\label{fulf}

\begin{theorem}\cite{fio}
\label{theo1}
If the noncommutative Hopf algebra $H$
is {\bf triangular} [\ie the
twist \F satisfies equation (\ref{cond2})], then,
setting $T\equiv\1\ot \1\equiv T'$ in eq. (\ref{def3}), 
$A^i,A_j^+$ close the quadratic commutation relations
\eqa
A^iA^+_j     & = & \delta^i_j\idA\pm R^{ui}_{jv}A^+_uA^v ,
\label{qccr1}\\
A^iA^j          & = & \pm R^{ij}_{vu}A^uA^v 
\label{qccr2}\\
A^+_iA^+_j & = & \pm R_{ij}^{vu}A^+_uA^+_v
\label{qccr3} 
\ena
where $R$ is the (numerical) quantum R-matrix of
$U\g$ in the representation $\rho$,
\eq
R^{ij}_{hk}:=\big[(\rho\ot \rho)(\R)\big]^{ij}_{hk}.
\en
\end{theorem}

\begin{theorem}\cite{fio}
\footnote{In the proof of theorem \ref{theo2} we made \cite{fio} 
essential use of the $U\g[[\hbar]]$-valued $2\times 2$ matrix
$(\rho\ot\id) \F$ found in Ref. \cite{zachos}.}
\label{theo2}
If $g=su(2)$ and $\rho\equiv$ fundamental representation,
it is possible to determine $T,T'$ such that
the elements $A^i,A^+_j\in\A[[\hbar]]$ ($i,j=\up,\downarrow$)
defined in formulae (\ref{def3}) are covariant under
$U_qsu(2)$ and satisfy the $U_qsu(2)$-covariant 
quadratic QCR \cite{puwo,pusz,wezu}
\eqa
  A^i  A^+_j     & = &\idA \delta^i_j\pm q^{\pm 1}R^{ui}_{jv}
  A^+_u  A^v ,
\label{qccr1n}\\
  A^i  A^j          & = & \pm q^{\mp 1}R^{ij}_{vu}  
A^u  A^v 
\label{qccr2n}\\
  A^+_i  A^+_j & = & \pm q^{\mp 1}R_{ij}^{vu}
  A^+_u  A^+_v,
\label{qccr3n} 
\ena
where 
$R$ is the $R$-matrix of 
$U_qsl(2)$. Moreover, $(A^i)^{\dagger}= A^+_i$ for the compact 
section 
$U_qsu(2)$ ($q\in\rn$). With this choice of $T,T'$, $A^+_i,A^j$ 
explicitly read, in the bosonic case,
\eq
\begin{array}{rclcrcl}
A^+_{\up} & = &\sqrt{(N_c^{\up})_{q^2}\over N_c^{\up}}
q^{N_c^{\downarrow}}A^{+c}_{\up} &\qquad
\qquad A^+_{\downarrow} & = &
\sqrt{(N_c^{\downarrow})_{q^2}\over N_c^{\downarrow}}
A^{+c}_{\downarrow}  \nonumber \\
A^{\up} & = &A_c^{\up}\sqrt{(N_c^{\up})_{q^2}\over N_c^{\up}}
q^{N_c^{\downarrow}} &\qquad
\qquad A^{\downarrow} & = &A_c^{\downarrow} 
\sqrt{(N_c^{\downarrow})_{q^2}\over N_c^{\downarrow}},
\end{array}
\label{lastb}
\en
and in the `fermionic' one 
\eq
\begin{array}{rclcrcl}
A^+_{\up} & = & q^{-N_c^{\downarrow}}A^{+c}_{\up} &\qquad
\qquad A^+_{\downarrow} & = &
A^{+c}_{\downarrow}  \nonumber \\
A^{\up} & = &A_c^{\up}q^{-N_c^{\downarrow}} &\qquad
\qquad A^{\downarrow} & = &A_c^{\downarrow},
\end{array}
\label{lastf}
\en
where
$N_c^{\up}:=A^{+c}_{\up}A^{\up}_c$, ~~
$N_c^{\downarrow}:=A^{+c}_{\downarrow}A_c^{\downarrow}$,~~
$(x)_{q^2}:={q^{2x}-1\over q^2-1}$.
\end{theorem}

\section{Application to QFT}
\label{new}

If the representation $\rho$ is reducible the algebra 
homomorphism $\sigma$ defined in eq. (\ref{jordan})
contains a sum over all the irreducible components. In
the case of a QFT, the generators of the Heisenberg
algebra are fields $\phi_c^{i_{\alpha}}(\vec{x})$ 
and there conjugate momenta $\pi^c_{i_{\alpha}}(\vec{x})$
(satisfying the commutation relations 
$[\phi_c^{i_{\alpha}}(\vec{x}),\pi^c_{j_{\beta}}
(\vec{x}')]_{\pm}=i\delta^{(3)}(\vec{x}-\vec{x}')\delta^
{\alpha}_{\beta}\delta^{i_{\alpha}}_{j_{\beta}}$), 
\ie  depend also on a continuous space index, so the sum 
entails also an integral:
\eq
\sigma(X^a)=i\int d^3x\x \sum\limits_{\alpha} 
\rho_{\alpha}(X^a)^{i_{\alpha}}_{j_{\alpha}}
\pi^c_{i_{\alpha}}(x)\phi_c^{j_{\alpha}}(x)
\qquad\qquad X^a\in\g;
\en
the index $\alpha$ enumerates all the kinds of different
fields (\ie particles) of the theory.  At the RHS we recognize
the charge $Q^a$ associated to the generator 
$X^a\in\g$.

The operators 
$\pi_{i_{\alpha}}(x)\phi^{j_{\alpha}}(x)$ which are obtained from
the canonical ones $\pi^c_{i_{\alpha}}(x)$, $\phi_c^{j_{\alpha}}(x)$
through the transformation (\ref{def3}), are
well-defined (nonlocal) composite operators on the Fock
space generated by $\phi_c^{i_{\alpha}}(\vec{x})$; they
act as $\pi^c_{i_{\alpha}}(x),\phi_c^{j_{\alpha}}(x)$ 
``dressed'' in a peculiar way by all the fields considered 
in the theory.  
In the case of a triangular 
Hopf algebra $H$, theorem \ref{theo1} implies that 
$\pi_{i_{\alpha}}(x),\phi^{j_{\alpha}}(x)$ satisfy the quadratic 
commutation relations
\eqa
\phi^{i_{\alpha}}(\vec{x})\pi_{j_{\beta}}(\vec{x}') &= 
&i\delta^{(3)}
(\vec{x}-\vec{x}')\delta^{\alpha}_{\beta}
\delta^{i_{\alpha}}_{j_{\beta}}
\pm R^{l_{\beta}i_{\alpha}}_{j_{\beta}m_{\alpha}}
\pi_{l_{\beta}}(\vec{x}') \phi^{m_{\alpha}}(\vec{x})\\
\phi^{i_{\alpha}}(\vec{x})\phi^{j_{\beta}}(\vec{x}') &= &
\pm R^{i_{\alpha}j_{\beta}}_{m_{\alpha}l_{\beta}}
\phi^{l_{\beta}}(\vec{x}') \phi^{m_{\alpha}}(\vec{x})\\
\pi_{i_{\alpha}}(\vec{x})\pi_{j_{\beta}}(\vec{x}') &= &
\pm R_{i_{\alpha}j_{\beta}}^{m_{\alpha}l_{\beta}}
\pi_{l_{\beta}}(\vec{x}') \pi_{m_{\alpha}}(\vec{x})
\ena
where 
$R^{i_{\alpha}j_{\beta}}_{m_{\alpha}l_{\beta}}=[(\rho_{\alpha}\ot 
\rho_{\beta})(\R)]^{i_{\alpha}j_{\beta}}_{m_{\alpha}l_{\beta}}$.
Because of remark 2 in sect. \ref{qcov}, in this case 
an invariant action
${\cal S}$ has the same functional dependence on 
$\pi_{i_{\alpha}}(x),\phi^{j_{\alpha}}(x)$ 
as on $\pi^c_{i_{\alpha}}(x),\phi_c^{j_{\alpha}}(x)$.

In the quasitriangular case $H=U_qsl(2)$ theorem
\ref{theo2} is not applicable, because in its present form
its validity is restricted only to the fundamental $\rho$ (so
the operators $\pi_{i_{\alpha}}(x),\phi^{j_{\alpha}}(x)$ do not 
satisfy quadratic commutation relations).
Whether some generalization of this theorem to arbitrary $\rho$ 
exists and these ideas can be applied to QFT also for 
quasitriangular $H$'s, is presently only matter of speculations.

\section*{Acknowledgments}
I am grateful to Prof.'s  V.\ K.\ Dobrev and  
H.-D.\ Doebner for their invitation to Group21.
It is a pleasure to thank  J. Wess for his support and
for the hospitality at his Institute.

\end{document}